 \documentclass[preprint2]{aastex}

\usepackage{epstopdf}

\slugcomment{To appear in The Astronomical Journal}

\shorttitle{TAROT early optical observations of GRBs}
\shortauthors{Klotz et al.}

\begin{document}


\title{
    Early optical observations of GRBs\\
    by the TAROT telescopes: period 2001-2008
}


\author{A. Klotz
\affil{Observatoire de Haute Provence (CNRS-OAMP),
    04870 Saint Michel l'Observatoire, France}
\affil{Centre d'Etude Spatiale des Rayonnements, 9 avenue colonel Roche,
    31028 Toulouse Cedex 4, France}
\email{klotz@cesr.fr}
}

\and

\author{M. Bo\"er
\affil{Observatoire de Haute Provence (CNRS-OAMP),
    04870 Saint Michel l'Observatoire, France}
\email{michel.boer@oamp.fr}
}
\and
\author{J.L. Atteia
\affil{Observatoire Midi Pyr\'en\'ees, LATT (UPS-CNRS), Universit\'e de Toulouse,
    31400 Toulouse, France}
\email{atteia@ast.obs-mip.fr}
}

\and
\author{B. Gendre
\affil{Laboratoire Astronomique de Marseille , LAM (UP-CNRS), Universit\'e de Provence,
    38, rue Fr\'ed\'eric Joliot-Curie, 13388 Marseille cedex 13, France}
\email{bruce.gendre@oamp.fr}
}




\begin{abstract}
The TAROT telescopes ({\it T\'elescopes \`a Action Rapide pour les Objets Transitoires})
are two robotic observatories designed to observe the  prompt optical emission
counterpart and the early afterglow of gamma ray bursts (GRBs). We present data acquired
between 2001 and 2008 and discuss the properties
of the optical emission of GRBs, noting various interesting results.
The optical emission observed during the prompt GRB phase is rarely
very bright: we estimate that 5\% to 20\% of GRBs exhibit
a bright optical flash (R\,$<$\,14) during the prompt gamma-ray emission,
and that more than 50\% of the GRBs have an optical
emission fainter than R\,$=$\,15.5 when the gamma-ray emission is active.
We study the apparent optical brightness distribution of GRBs
at 1000\,sec showing that our observations confirm the distribution 
derived by other groups.
The combination of these results with those obtained by other rapid 
slewing telescopes allows us to better characterize the early optical emission of GRBs and
to emphasize the importance of very early multi-wavelength GRB studies for the
understanding of the physics of the ejecta.
\end{abstract}



\keywords{telescope: general ---
GRBs: individual (
\objectname{GRB~010213},
\objectname{GRB~020531},
\objectname{GRB~030115},
\objectname{GRB~030324},
\objectname{GRB~030329},
\objectname{GRB~030501},
\objectname{GRB~040916},
\objectname{GRB~041006},
\objectname{GRB~041218},
\objectname{GRB~041219A},
\objectname{GRB~050209},
\objectname{GRB~050306},
\objectname{GRB~050408},
\objectname{GRB~050416},
\objectname{GRB~050504},
\objectname{GRB~050505},
\objectname{GRB~050520},
\objectname{GRB~050525A},
\objectname{GRB~050730},
\objectname{GRB~050803},
\objectname{GRB~050824},
\objectname{GRB~050904},
\objectname{GRB~051211A},
\objectname{GRB~051211B},
\objectname{GRB~051221B},
\objectname{GRB~060111A},
\objectname{GRB~060111B},
\objectname{GRB~060124},
\objectname{GRB~060512},
\objectname{GRB~060515},
\objectname{GRB~060814},
\objectname{GRB~060904A},
\objectname{GRB~060904B},
\objectname{GRB~060919},
\objectname{GRB~061019},
\objectname{GRB~061027},
\objectname{GRB~061028},
\objectname{GRB~061110B},
\objectname{GRB~061217},
\objectname{GRB~061218},
\objectname{GRB~061222B},
\objectname{GRB~070103},
\objectname{GRB~070227},
\objectname{GRB~070330},
\objectname{GRB~070411},
\objectname{GRB~070412},
\objectname{GRB~070420},
\objectname{GRB~070508},
\objectname{GRB~070521},
\objectname{GRB~070621},
\objectname{GRB~070628},
\objectname{GRB~070913},
\objectname{GRB~070920A},
\objectname{GRB~070920B},
\objectname{GRB~071010A},
\objectname{GRB~071101},
\objectname{GRB~071109},
\objectname{GRB~071112B},
\objectname{GRB~071112C},
\objectname{GRB~080129},
\objectname{GRB~080207},
\objectname{GRB~080210},
\objectname{GRB~080315},
\objectname{GRB~080330},
\objectname{GRB~080413A},
\objectname{GRB~080430},
\objectname{GRB~080516},
\objectname{GRB~080603B},
\objectname{GRB~080903}
) ---
}


\section{Introduction}

The discovery of Gamma-Ray Burst (GRB) optical afterglows in 1997 \citep{vanParadijs1997}
and the creation of the Internet message distribution service by BACODINE and
the Gamma-ray burst Coordinate Network \citep[GCN, ][]{Barthelmy1998}
gave the impulse to look for the \textit{prompt} optical emission of GRBs\footnote{Throughout this paper the term 'prompt emission' refers to the photons detected when the 
GRB is {\it active} (or in the {\it prompt} phase). This {\it prompt} phase is
arbitrarily defined as starting with the trigger and ending after $T_{90}$ seconds, 
where $T_{90}$ is the time during which 90\% of the burst photons are detected
(from 5\% to 95\% of the cumulated photon fluence).}.
The first detection of the prompt optical emission from a GRB was performed
by ROTSE-I on GRB~990123~\citep{akerlof99}.
The exceptionally bright optical emission of GRB~990123 (V$\,\sim\,$9) 
was uncorrelated with the gamma-ray light-curve.
After 2001, the operation of satellites providing fast GRB localizations (HETE-2,
INTEGRAL, \emph{Swift}, AGILE, \emph{Fermi}) have allowed the detection of some GRBs 
with \textit{bright} optical emission during their gamma activity (typically R\,$<$\,13).
These bursts usually follow a behavior similar to GRB~990123, with a single bright 
peak at optical wavelengths \citep[e.g. GRB~060111B][]{Klotz2006}.

On the other hand, using the trigger delivered by the
\emph{Swift} satellite \citep{Gehrels2004},
\citet{Vestrand2006} detected a faint optical emission
correlated with the gamma-ray light-curve in GRB~050820A.
Though there are few bursts that are well observed at optical wavelengths
during the gamma-ray emission, it seems that GRB~990123
and GRB~050820A are prototypes of two categories
of GRBs: {\it i)} those with bright optical transients
uncorrelated with the gamma-ray activity often interpreted as the result of
reverse shocks \citep[]{Jin2007}, and
{\it ii)} those with faint optical emission correlated with
the gamma-ray light-curve often interpreted as the low energy tail
of the high-energy emission \citep{Genet2007}.

Understanding the multi-wavelength emission of GRBs during the 
prompt phase is very important, as it provides direct insight into the
composition of the ejecta and its physical conditions. 
In the optical such studies require the combination of GRB satellites
providing fast alerts and rapid slewing telescopes on the ground.
The launch of \emph{Swift} has increased by an order of magnitude the number
of alerts sent to ground observatories, allowing small robotic telescopes
to contribute actively to the study of the prompt GRB emission at optical
wavelength. This is in particular the case of the two TAROT robotic telescopes,
which are designed to observe the optical emission during the first seconds 
following a GRB alert \citep{Klotz2009}. They are able to
slew very rapidly to any sky location above their horizon upon receipt of GCN notices.

This paper is devoted to the analysis of the GRB
observations made by TAROT during the period 2001-2008.
Depending on the burst, observations started either during the prompt emission or
the afterglow. Section \ref{Observatories} summarizes the main
technical characteristics of the two robotic TAROT observatories.
Section \ref{Observations} describes the acquisition strategies
and image processing methods.
Section \ref{prompt} is devoted to the analysis of the prompt
optical emission.
From our data we derive the apparent optical brightness distribution 
of GRBs at 1000\,sec in Section \ref{afterglow}.
In  Section \ref{Standard candles} we discuss
how the afterglows seen by TAROT can be used  for cosmological studies.

\section{TAROT observatories}
\label{Observatories}

The TAROT telescopes were designed in 1995, two years before
the first optical detection of a GRB. At this epoch, the
size of the error boxes provided by the CGRO-BATSE experiment
\citep{Fishman1992} was about $5^{\circ}\times 5^{\circ}$.
As a consequence, the field of view (hereafter FoV) of TAROT
was designed to be large enough to
cover as much as possible of the BATSE error boxes.
Two telescopes were planned: one in France (operating since 1998)
at the Calern observatory (Observatoire de la C\^ote d'Azur, CNRS), 
the second in Chile (active since 2006) at the La Silla European Southern
Observatory. 
In order to reach
the position of the GRB provided by the GCN as
quickly as possible, the mount can move as fast as
50$^{\circ}$\,sec$^{-1}$. Primary mirrors are 25\,cm in diameter
and 85\,cm in focal length. These observatories
are fully autonomous and robotized (scheduling, acquisition
and data management).
A detailed description of the telescopes
can be found in \citet{Klotz2009}. 

From 1998 to 2000, the detector was an APOGEE camera based on a Kaf-1300
CCD chip that covered only $1.3^{\circ}\times 1.1^{\circ}$.
The slewing strategy was to make a mosaic of 25
images in order to cover most of the BATSE FoV.
Since the exposure time was 30\,sec and the readout time
was 30\,sec, it implied that the complete survey lasted
about 25 minutes for a limiting magnitude of 14.
Twenty-one BATSE triggered GRBs were observed by TAROT 
during that period \citep{Boer2001}.
Due to the large delays between alert and observations
and the poor limiting magnitude,
we do not include these events  in the present analysis.

After 2001, the HETE \citep{Ricker2001} error
boxes were smaller than $\simeq 20 \arcmin$. 
They fitted easily in one $1.9^{\circ}\times 1.9^{\circ}$ image made by the
new custom camera based on a TI-7899 chip from Thomson, allowing
rapid slewing to the source location and an earlier observation 
of the optical emission.
In 2003, this camera was replaced by an
ANDOR 436, based on a Marconi-4240 back illuminated CCD.
The FoV remains essentially the same: $1.86^{\circ}\times 1.86^{\circ}$.
Since 2003, TAROT observed GRBs triggered by
INTEGRAL and by  \emph{Swift} \citep{Gehrels2004} with error boxes smaller than 3$\arcmin$.

The last significant evolution of the program was the installation of 
TAROT `La Silla' in Chile whose first light occurred in 2006 with
the same ANDOR 436 camera, and characteristics similar to TAROT `Calern'.

\begin{figure}[htb]
\begin{center}
\begin{tabular}{c}
\includegraphics[width=1\columnwidth]{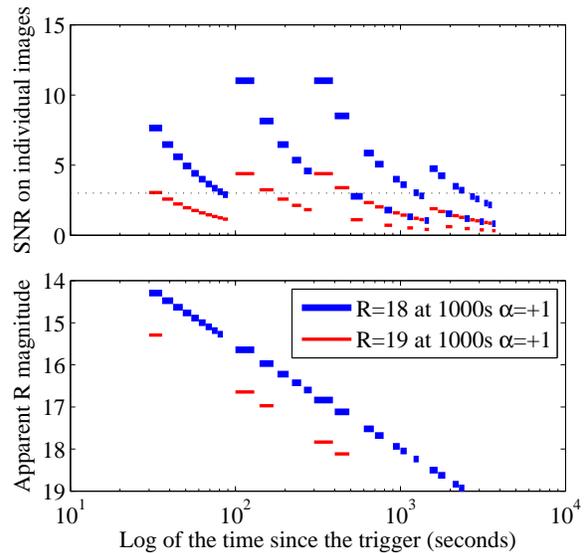}
\end{tabular}
\caption{
The observing strategy is illustrated by two simulated
afterglows: R\,$=$\,18 at 1000 sec with $\alpha=+1$ ($f(t) \varpropto t^{-\alpha}$, thick blue lines
in the online edition), and
R\,$=$\,19 at 1000 sec with $\alpha=+1$ (thin red lines in the online edition)
Top panel: Signal to noise ratio (SNR) as a function of elapsed time since trigger considering the
actual series  of exposure performed by TAROT (see
section \ref{Observations}). Bottom panel: the
corresponding time resolved light-curves
from individual images. For the fainter GRB,
the final light-curve is made by co-adding
individual images in order to increase the detectivity.
}
\label{tarot_strategy}
\end{center}
\end{figure}

\section{GRB observations with TAROT}
\label{Observations}

Since December 2005 we have adopted a strategy resulting in the
acquisition of a series of exposures that optimizes the signal
to noise ratio (SNR) and the time resolution. The first
image of this series is an unfiltered 60-sec exposure
taken with a slight drift along the diurnal
motion. This strategy results in a trailed image of the 
afterglow with no dead-time. Such a trailed image will readily display
any rapid variations of the optical emission.
The drift is calibrated to get 6\,sec\,pixel$^{-1}$. 
For the brightest afterglows, this provides a time-resolved light-curve with a temporal 
resolution of several seconds (typically 5\,--\,10).
After the trailed image, five 30-sec unfiltered exposures are taken
with no drift. This then completes the first series of images.
After this series, four series of three 90-sec long images are made,
the first two images of each series are unfiltered and the third is taken with an R filter.
When possible, 180-sec exposures
are taken with the same filter sequence until 4 hours after the trigger.
The resulting SNR and the light-curve are
presented for two simulated cases in Figure~\ref{tarot_strategy}.
A late follow-up
is performed, acquiring a series of VRI 180-sec
images every 15\,minutes until the end of the night.

\begin{figure*}[htb]
\begin{center}
\begin{tabular}{c}
\includegraphics[width=1.5\columnwidth]{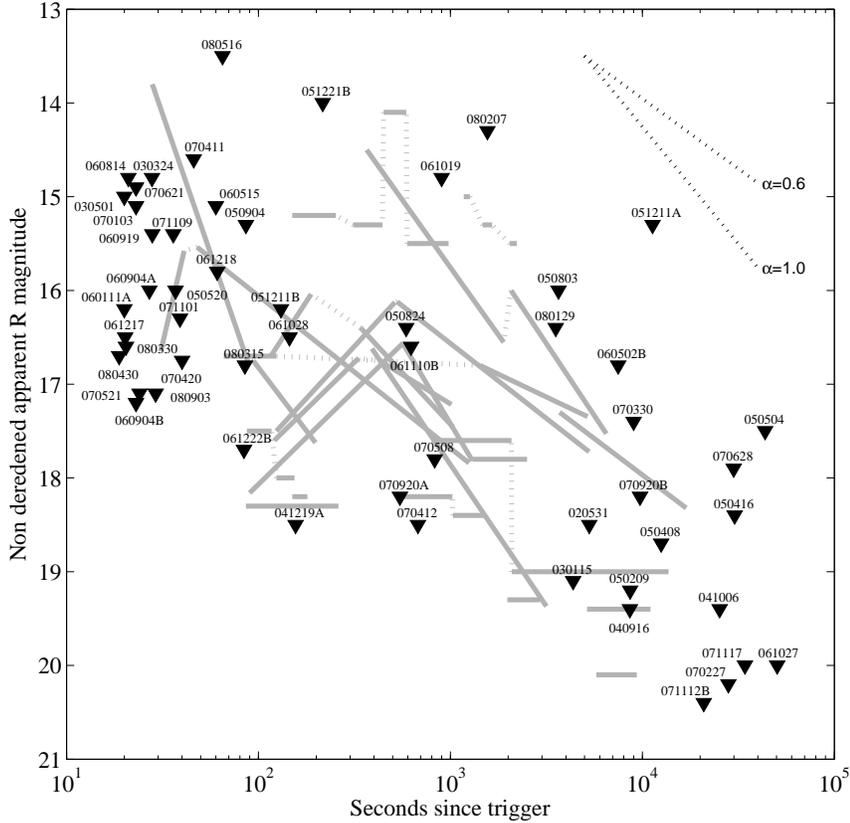}
\end{tabular}
\caption{
TAROT observations of the early optical emission from GRBs: 
the triangles represent upper limits for the first time bin and the 
thick gray lines are the light-curves of detected sources.
Upper limits are identified with the name of the GRB, for light-curve identification,
see Fig~\ref{tarot_grid1}.
The dotted lines on the upper right corner illustrate two typical afterglow decay slopes.
The data are from Tables~\ref{tarotgrbs1} and \ref{tarotswift} of this paper.
}
\label{tarot_firsts}
\end{center}
\end{figure*}

\begin{figure*}[htb]
\begin{center}
\begin{tabular}{c}
\includegraphics[width=1.5\columnwidth]{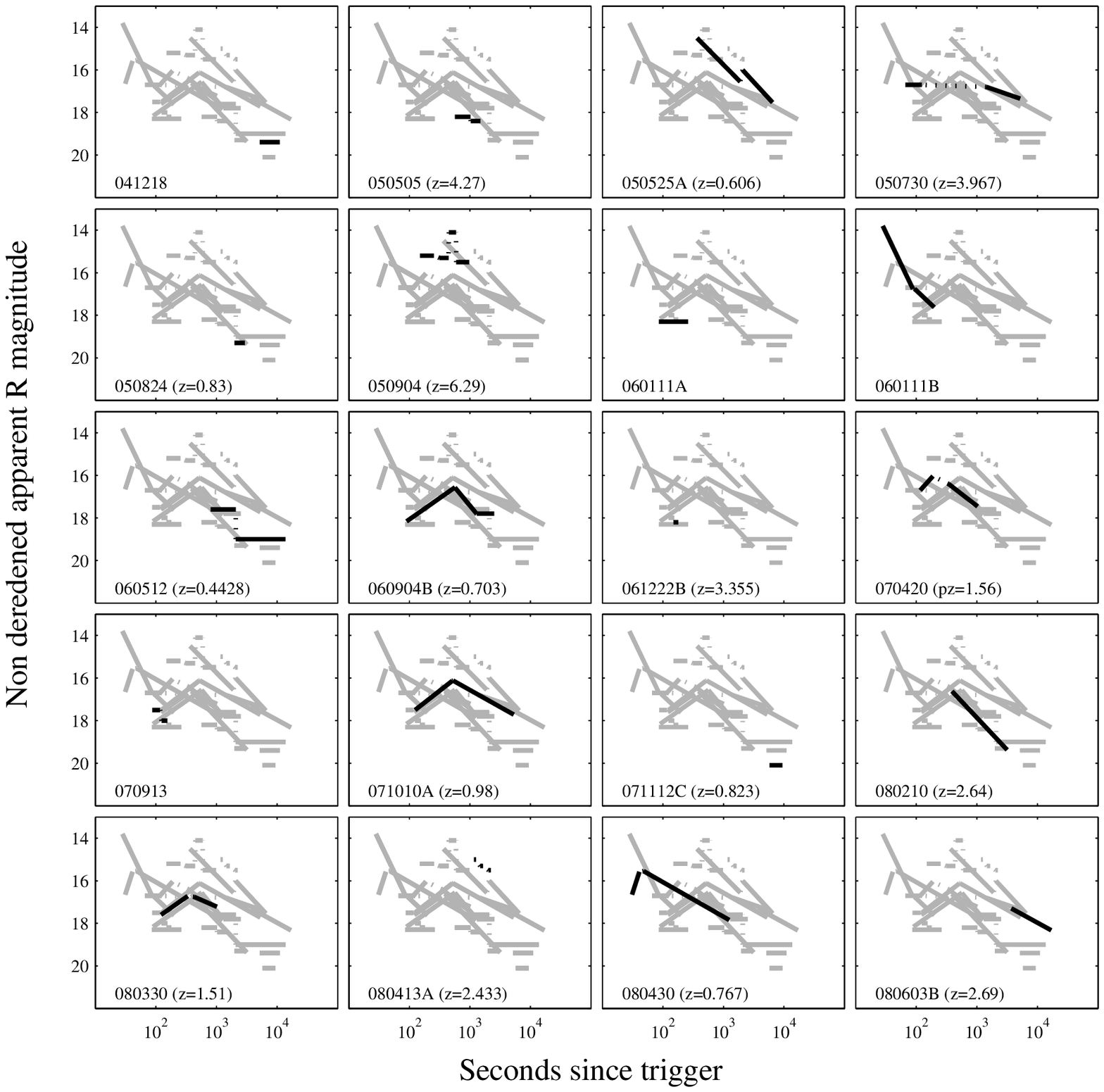}
\end{tabular}
\caption{
TAROT detections of GRB afterglows. Twenty GRB afterglows 
detected by TAROT are shown for comparison, one per box.
The black curves correspond to the GRB indicated at the bottom 
of the box. Gray lines are the other GRBs detected by TAROT.
16 GRBs have redshifts -- indicated at the bottom of their respective
boxes above. (Note that the `pz' for GRB 070420 stands for `pseudo-redshift'.)
}
\label{tarot_grid1}
\end{center}
\end{figure*}

Tables~\ref{tarotgrbs1} and \ref{tarotswift} 
summarize the TAROT optical observations of 72~GRBs that were observed
between 2001 and 2008. During that period, twenty optical
transients were detected. For eleven of them there is more than one detection available,
allowing the measure of the temporal decay index.
When the gamma-ray emission is still active,
fourteen GRBs were observed and
3 of them showed an optical emission
detected by TAROT.

The majority of data displayed in Tables~\ref{tarotgrbs1}
and \ref{tarotswift} have been published in GCN Circulars
and various papers (listed in column 6 of each table).
We correct here some errors published in the GCN circulars for
GRB~050520 and GRB~050730, due to bad dates and poor
quality calibration, now refined.
The limiting magnitude of the first image of GRB~030501
was measured during the prompt emission.
GRB~060124 was embeded in a bright star. The optical
counterpart was extracted by image difference
that leads to a lower limit.
We also performed a full analysis of the images
of GRBs for which TAROT started observing very late:
GRB~050408, GRB~050416, GRB~050504 and GRB~050824.
These results are summarized in Figure \ref{tarot_firsts} which displays 
the detections and one upper limit for those GRBs which have not been detected.
Figure \ref{tarot_grid1} displays all the detected optical
emissions. On these figures, we have plotted the observed
magnitudes, without any correction for the galactic extinction
or for the cosmological effects.

The afterglow decay, which can start during the prompt event,
is usually described by a power-law of index
 $\alpha$ ($f(t) \varpropto t^{-\alpha}$, where $f(t)$ is the observed flux).
According to \citep{Zeh2006}, $\alpha$ lies between 0.5 and 1.7 with a median
value of 1.0.
Additional features can make the afterglow decay more complex.
Many afterglows detected by TAROT exhibit an increase in brightness
until a few hundreds of seconds (GRB~050904,
GRB~060904B, GRB~070420, GRB~071010A, GRB~080330, GRB~080430).
In all cases, the decay regime of the afterglow, $t^{-\alpha}$, is well
established after 600\,sec. In the cases of GRB~050525A and 
GRB~060904B, the light-curves show a plateau or a rebrightening near
2$\times$10$^3$\,sec in the rest frame.

\section{Optical emission during gamma-ray activity}
\label{prompt}

\begin{figure}[htb]
\begin{tabular}{c}
\includegraphics[width=0.99\columnwidth]{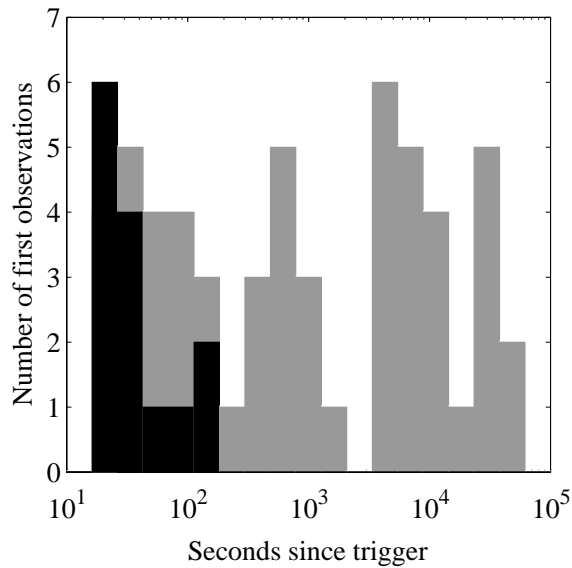}
\end{tabular}
\caption{
Histogram of the observation delays between the trigger
and the start of the first TAROT exposure.
Black bars represent exposures which started while the
gamma-ray emission was still active.
}
\label{tarot_histofirsts}
\end{figure}

Because the TAROT slewing time is about 5\,sec, 
and the \emph{Swift}\,-\,GCN notification time takes about 20\,sec,
only the longest GRBs can be studied during their prompt emission.
Figure \ref{tarot_histofirsts} shows the
histogram of the delay between the trigger and the start of the first observation.
The black bars represent events that were observed by TAROT before the end 
of the prompt emission, there are 14 such events. 
This figure shows that 90\,\% of the bursts (10/11) 
are still in the prompt phase after 40\,sec and that no GRB afterglow
is observed during the prompt emission after a delay of about 200\,sec.

\begin{figure}[htb]
\begin{tabular}{c}
\includegraphics[width=0.99\columnwidth]{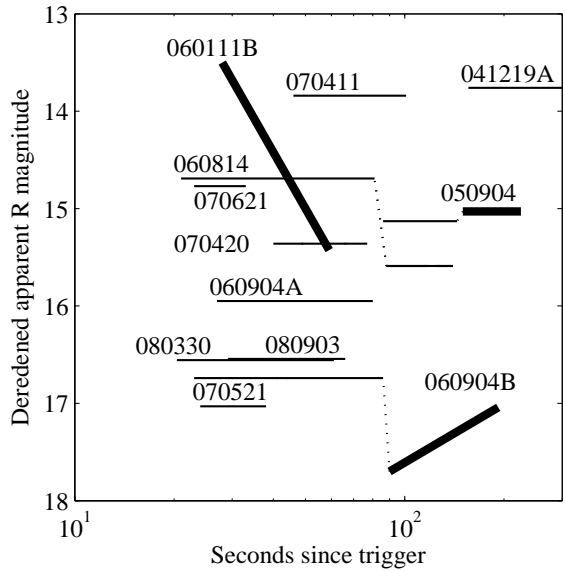}
\end{tabular}
\caption{
Plot of TAROT measurements performed
while the GRB was still active. Thick lines are
detections and horizontal thin lines represent upper limits. 
The magnitudes are de-reddened from the galactic extinction. 
GRB~030501 and GRB~050306 are not shown on this plot because
of the poor constraints on the brightness of their optical afterglows
resulting from a high galactic absorption (GRB~030501) 
and/or a poor sensitivity limit (GRB~050306).
}
\label{tarot_prompts}
\end{figure}

Figure \ref{tarot_prompts}
displays the measurements
done by TAROT during the prompt phase of 12 GRBs.
Though 14 events were observed by the TAROTs during the prompt
emission, only 12 of them appear
in Figure \ref{tarot_prompts} because we have eliminated
two bursts.
GRB~030501 was observed through a high extinction region 
of the Galaxy, preventing us from giving meaningful constraints on the brightness
of its afterglow and GRB~050306 has a
poor upper magnitude limit due to a low elevation
and the presence of haze.
Figure \ref{tarot_prompts} shows that GRB~060111B was the brightest 
of the detected emissions, and it exhibits a fast decay.
The rise of the afterglows of GRB~050904 and GRB~060904B
started before the end of the gamma-ray activity
but no emission was detected during the first
trailed exposure.
This figure also shows that more than 50\% 
of the GRBs are fainter than R\,$=$\,15.5 during the prompt phase.
Optically bright GRBs with a rapid decay phase, like GRB~060111B (or GRB~990123) 
seem to be rare: at most 5 to 20\% of the bursts.

\section{Afterglow brightness distribution}
\label{afterglow}

Several authors have discussed the afterglow apparent brightness distribution
\citep[{\it e.g.} ][]{Berger2005,Fiore2006}.
The first problem is to define the
brightness since the source luminosity decreases rapidly.
Here we follow the approach of \citet[][hereafter AS07]{Akerlof2007}, 
and we define the reference brightness
as the apparent R magnitude 1000\,sec after the GRB trigger,
deredened from the galactic extinction.
This enables us to alleviate somewhat the biases introduced by the
erratic behaviour of the optical emission which is sometimes observed 
during the first hundreds of seconds after the trigger.

When TAROT observations stop before 1000\,sec
we extrapolate the light-curve to 1000\,sec
using a temporal index $\alpha=+1.0$.
For observations starting between 1000 and 2000\,sec,
we extrapolate the light-curve backward with the same index.
We do not include GRBs for which the
first exposure  started after 2000\,sec since the extrapolation
may lead to additional biases.
In total we have
45 events with 12 positive measurements
and 33 upper limits. This is much less than the
109 GRBs with 43 positive detections quoted in AS07. However, AS07 mixes data from 29
different telescopes from 6 to 820\,cm aperture and
$\sim$40\% 
of data come from the UVOT onboard the \emph{Swift} satellite, 
which may introduce various biases. In particular, unlike ground telescopes 
the UVOT cannot observe in the R band, requiring a k-correction 
based on the optical spectral index and the host extinction 
which is usually poorly constrained.
Our study has only five GRBs in common with AS07.
The TAROT study completes what was done by AS07
with a more homogeneous data set. However, the small
TAROT aperture induces a brighter cut-off of the
distribution.

\begin{figure}[htb]
\begin{tabular}{c}
\includegraphics[width=0.99\columnwidth]{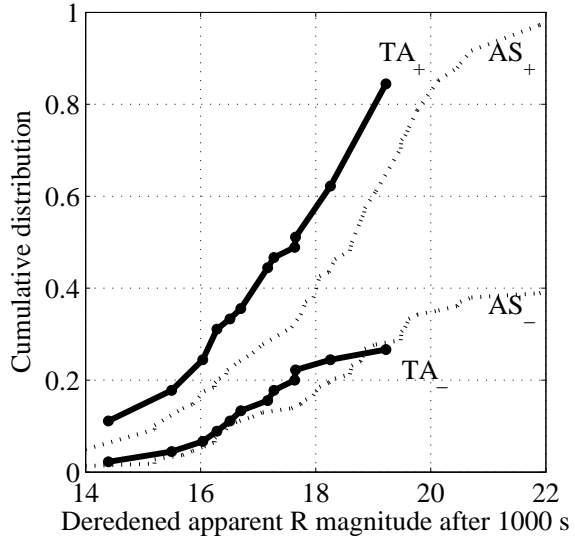}
\end{tabular}
\caption{
Apparent optical brightness distribution of afterglows detected 1000\,sec after
the GRB. The upper thick line (TA$_{+}$) is constructed assuming that
undetected afterglows are just below ({\it i.e.}, dimmer than) the limiting
magnitude. It corresponds to the upper value of the
actual distribution (optimistic case in brightness).
The lower thick line (TA$_{-}$) is constructed considering
all afterglows which are not detected are fainter than
the faintest detected afterglow. 
Dashed lines (AS$_{-}$ and AS$_{+}$)
are the same limits computed from Tables 1 \& 2 of \citet{Akerlof2007}.
}
\label{tarot_magtref}
\end{figure}

Figure~\ref{tarot_magtref} compares the cumulative
distributions of AS07 (AS) and TAROT (TA)
at 1000\,sec.
The most optimistic cases (AS$_{+}$
and TA$_{+}$) are computed with the assumption that the non-detections 
are just below the observed limiting magnitude.
This is an extreme case but it allows us to
place an upper limit on the distribution.
The relative position of TA$_{+}$ above AS$_{+}$ reveals the differences in
limiting magnitudes of the two studies.
We calculated also the most pessimistic cases (AS$_{-}$
and TA$_{-}$) assuming that the non-detections
are fainter than the faintest detected afterglow.
The TA$_{-}$ distribution follows very well the
AS$_{-}$ one, until R\,$=$\,17.5 where TA$_{-}$ becomes
almost flat. One can notice that AS$_{-}$ also becomes also
flat for R\,$>$\,20. These two magnitudes corresponds
to the typical limiting magnitudes of TAROT and UVOT
respectively.
The actual distribution must lie between the two + and - curves.
AS07 used a recursive method to correct the
AS$_{-}$ data in order to get the figure 4
of their paper. However, their Figure 4 curve is very
close to AS$_{-}$. 

The TAROT study confirms the apparent brightness distribution obtained by AS07
for R\,$<$\,17.5, minimizing the telescope
bias. We can see that the proportion of afterglows brighter than
R\,$=$\,14 at 1000 sec is very small (a few percent at most).
The slope of the distribution increases
slowly until R\,$=$\,20. The flatness of AS$_{-}$ and TA$_{-}$ curves
may be due to instrumental limitations (i.e. the limiting magnitude)
but it can also be the result of the
high redshift GRBs ($z$\,$>$\,5) that are no longer
detectable in the R band or by dark GRBs
embedded in a dust cocoon that extincts optical
wavelengths. Anyway, TAROT confirms that at least 10\%
of the afterglows are brighter than R\,$=$\,16.5,
1000\,sec after the trigger.

\begin{figure}[htb]
\begin{center}
\begin{tabular}{c}
\includegraphics[width=\columnwidth]{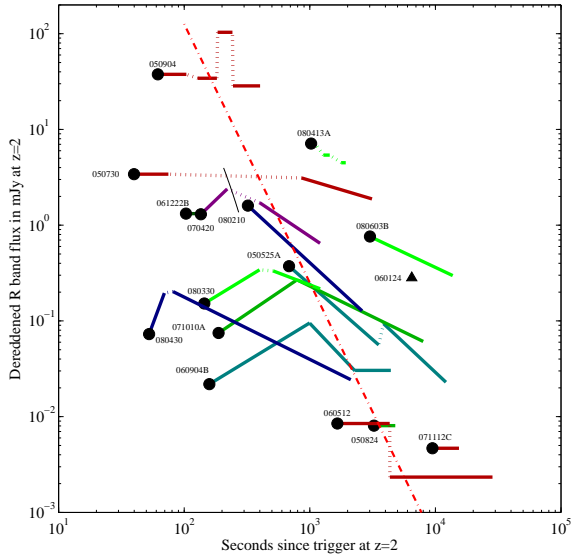}
\end{tabular}
\caption{
Light-curves of afterglows with known 
redshifts or pseudo-redshifts, placed at a common redshift of $z$\,=\,2.
The dash-dot line is defined as the best correlation
between peak times and fluxes for eleven fast rising
afterglows \citep{Panaitescu2008}. The peak of the emission
of slow rising afterglows appear less correlated with the maximum
of the flux than in PV08.
This figure appears in colors in the electronic version of this paper.
}
\end{center}
\label{tarot_firstzs}
\end{figure}

\section{Rising afterglows as standard candles ?}
\label{Standard candles}

Early GRB afterglows can shine orders of
magnitudes brighter than quasars, enabling them to be seen
across the entire Universe and opening the possibility to use them 
to study the geometry of the Universe.
However, using GRBs to probe the cosmology
can be powerful only if some of their properties
are constant and independent from burst to burst.
\citet[][hereafter PV08]{Panaitescu2008} used 11 fast rising afterglows
to derive a relation
between $t_p$, the time of their peak brightness 
and $F_p$, the brightness at their peak (in the GRB rest-frame).
Figure\,7 
displays the optical light-curves 
observed by TAROT, for GRBs having a spectroscopic redshift or a
pseudo-redshift \citep[e.g ][]{Atteia2005}.
The axes of Figure\,7 
are the same as those of Figure 2 of PV08 and
the dash-dot line is the PV08 $t_p-F_p$ relation.
We used a flat cosmology with $\Omega_m = 0.27$,
$\Omega_{\Lambda} = 0.73$, and h$_0 = 0.71$ to
scale fluxes at  $z$\,$=$\,2. TAROT observations of
GRB~050904 and GRB~060904B have been used by PV08
in their study. GRB~060904B and GRB~071010A lie just below the
PV08 relation. There is only a pseudo redshift
for GRB~070420. 
The black line passing through the light-curve
of GRB~070420 indicates the confidence region
of the location of the maximum of brightness.

It must be mentioned that
GRB~060904B, GRB~070420, GRB~071010A and GRB~080330 are slow
rising afterglows according to the definition of PV08. These
afterglows have fainter fluxes 
in comparison to the PV08 $t_p-F_p$ relation.
GRB~080430 is clearly in the category of fast
rising afterglows but is very far from the PV08 $t_p-F_p$ relation.
At present, the large dispersion of these values does not
allow us to consider slow rising afterglows as useful
candles for cosmology. The case of GRB~080430 may
indicate that the PV08 $t_p-F_p$ relation is
not valid for faint fast rising afterglows (say $<1$ mJy at $z$\,$=$\,2).


\section{Conclusion}

Despite its small aperture, TAROT participates
actively in the observation of the early
optical emission of GRBs. A few GRBs have allowed TAROT 
to make original contributions to the field of GRB studies:
GRB~050904 (optical emission during the prompt at $z$\,=\,$6.3$),
GRB~060111B (time-resolved optical emission during
the prompt phase), GRB~060904B, GRB~070420, GRB~071010A,
GRB~080330 and GRB~080430 (five rising afterglows).
We argue that more than half of GRBs have an early optical
emission which is fainter than R\,$=$\,15.5 when the gamma emission is still active
and we find that only 5 to 20\% of the bursts produce a
bright optical flash, confirming UVOT observations \citep{Roming2006}.
However, these rare bursts are especially interesting and spectacular when
they are detected (e.g. GRB~060111B and the recent detection of the optical flash 
associated with GRB~080319B \citep{Cwiok2008}).
The slope of the afterglow apparent brightness distribution provides new observational data
that can be explained by the conjunction of intrinsic
parameters (the GRB luminosity function), and extrinsic parameters 
(the redshift distribution of GRBs).
We confirm the result of AS07 that 90\%
of the afterglows are fainter than R\,$=$\,16.5, 1000\,sec
after the trigger.
We show that a significant fraction of afterglows display an
increase in brightness during the first few hundreds of
seconds after the trigger.
These statistical results enable us to size the
next generation of robotic observatories (see AS07 for a discussion).
They will feature 1-meter class telescopes and near infrared detectors. 
Prototypes of these evolutions are 
REM \citep{Conconi2004},
BOOTES-IR \citep{Castro-Tirado2005},
PAIRITEL \citep{Bloom2006}, while GROND \citep{Greiner2008}
already represents the new generation of powerful facilities dedicated 
to the follow-up of GRB early afterglows. 
Finally, the bulk of the results presented in this paper rest on
the large number of GRB localizations provided by \emph{Swift} and on the
flawless operation of the TAROT robotic observatories. 
These conditions will continue to be realized in the near future with the
expectation, thus, of many more interesting discoveries in the coming years.


\acknowledgments{
Acknowledgments:
The TAROT telescope has been funded by the {\it Centre National
de la Recherche Scientifique} (CNRS), {\it Institut National des
Sciences de l'Univers} (INSU) and the Carlsberg Fundation. It has
been built with the support of the {\it Division Technique} of
INSU. We thank the technical staff contributing to the TAROT project:
A. Abchiche, G. Buchholtz, A. Laloge, A. Mayet, M. Merzougui, A.M. Moly,
S. Peruchot, H. Pinna, C. Pollas, P. \& Y. Richaud, F. Vachier, A. le Van Suu.
Bruce Gendre acknowledges a post-doctoral grant funded by the Centre
National d'Etudes Spatiales.
We thank Alan Imerito for interesting comments on the paper.
}

\clearpage

\begin{table*}[htb]
\caption{
TAROT measurements of GRB optical transients between 2001\,--\,2004.
Timings are calculated from the GRB trigger.
An asterisk ($^*$) indicates that the measurement was done when the GRB was still active.
CR magnitude means Clear filter calibrated with R magnitudes of USNO-B1 stars.
}
{\scriptsize
\begin{center}
\begin{tabular}{c c c c c c c r}
\hline\hline
GRB    & Trigger & T$_{90}$ & TAROT          & Magnitude  & Reference & Redshift & A$_R$\cr
       &         &          & start - end &                       & & &\cr
\hline
010213 & HETE & 30s & 37.2 - 37.4 h & CR$>$17 & GCNC 936 & & 0.09\cr
020531 & HETE & $<$1s & 1.47 - 1.57 h & CR$>$18.5 & \citet{Klotz2003}  & & 0.31\cr
030115 & HETE & 20s  & 1.21 - 1.56 h & BR$>$19.1 & GCNC 1810  & 2.20 & 0.05\cr
030324 & HETE & 8.7s  & 27.9 - 57.9 s & CR$>$14.8 & GCNC 1961  & & 0.07\cr
030329 & HETE &  25s  & 6.37 - 6.37 day & CR$=$18.55 $\pm$ 0.36 & GCNC 2247  & 0.16854 & 0.07\cr
       &      &       & 10.3 - 10.3 day & CR$=$18.81 $\pm$ 0.44 & GCNC 2247  & 0.16854 & 0.07\cr
030501 & INTEGRAL & 40s & 20 - 50 s$^*$ & CR$>$15 & This study & & 39.55\cr
       &          &      & 20 - 1020 s & CR$>$18 & GCNC 2224  & & 39.55\cr
040916 & HETE & 450s & 2.39 - 2.89 h & CR$>$19.4 & GCNC 2729  & & 0.16\cr
041006 & HETE & 20s & 7.02 - 8.18 h & CR$>$19.4 & GCNC 2784  & 0.716 & 0.06\cr
041218 & INTEGRAL & 60s & 1.43 - 3.06 h & CR=19.4 $\pm$ 0.2 & GCNC 2904  & & 1.68\cr
041219A & INTEGRAL & 520s & 2.6 - 11.2 min$^*$ & CR$>$18.5 & GCNC 2905  & & 4.74\cr
\hline
\end{tabular}
\label{tarotgrbs1}
\end{center}
}
\end{table*}

\begin{deluxetable}{c c c c c c c c r}
\tabletypesize{\footnotesize}
\tablecaption{TAROT measurements of GRB's optical transients in 2005\,--\,2007.
Dates are calculated from the GRB trigger.
An asterisk ($^*$) indicates that the measurement was done when the GRB was still active.
The right arrow symbol ($\rightarrow$) designates a series of measurements with the
magnitude being that at the starting time followed by
temporal decay according to the $f(t) \varpropto t^{-\alpha}$ law.
Redshift written in italic means that they are pseudo-redshifts \citep{Atteia2005}. Site of observation is Calern Observatory (c) or La Silla Observatory (s).\label{tarotswift}
CR magnitude means Clear filter calibrated with R magnitudes of USNO-B1 stars. CI means
clear filter calibrated with I magnitudes. Note that all magnitudes displayed are not corrected
for galactic extinction.
}
\tablehead{\colhead{GRB } &\colhead{Trigger} &\colhead{T$_{90}$ } &\colhead{TAROT} &\colhead{Magnitude} &\colhead{Reference} &\colhead{Site} &\colhead{Redshift} &\colhead{A$_R$}\\
\colhead{}  &\colhead{} &\colhead{}  &\colhead{start - end} & \colhead{}   &\colhead{} &\colhead{}   &\colhead{} &\colhead{}}
\startdata
050209 & HETE & 40s & 2.40 - 3.22 h & CR$>$19.2 & GCNC 3020 & c &{\it 2.93 $\pm$ 1.6} & 0.11\\
050306 & SWIFT & 180s  & 126 - 157 s$^*$ & CR$>$11 & GCNC 3084  & c & & 1.814\\
       &       &        & 210 - 440 s & CR$>$13 & GCNC 3084  & c & & 1.814\\
       &       &        & 505 - 619 s & CR$>$14.0 & GCNC 3084  & c & & 1.814\\
       &       &        & 45 - 80 min & CR$>$17.5 & GCNC 3084  & c & & 1.814\\
050408 & HETE & 34s  & 3.48 - 3.84 h & CR$>$18.7 & This study & c & 1.2357 & 0.07\\
050416 & SWIFT & 2.4s & 8.40 - 9.52 h & CR$>$18.4 & This study  & c & 0.6535 & 0.08\\
050504 & INTEGRAL & 80s & 12.11 - 12.15 h & CR$>$17.5 & This study  & c & & 0.03\\
050505 & SWIFT & 63s & 9.46 - 17.06 min & CI=18.2 $\pm$ 0.2  & GCNC 3403  & c & 4.27 & 0.06\\
       &       &     & 17.18 - 24.78 min & CI=18.4 $\pm$ 0.2  & GCNC 3403  & c & 4.27 & 0.06\\
       &       &     & 27.57 - 37.30 min & CI$>$18.8  & GCNC 3403  & c & 4.27 & 0.06\\
050520 & INTEGRAL & 20s & 37 - 52 s & CR$>$16.0 & This study  & c & & 0.04\\
       &          &     & 59 - 74 s & CR$>$16.0 & This study  & c & & 0.04\\
       &          &     & 80 - 441 s & CR$>$17.2 & This study  & c & & 0.04\\
050525A & SWIFT & 8.8s & 6.1 $\rightarrow$ 31.6 min & CR=14.5 $\alpha$=1.15 & \citet{Klotz2005} & c & 0.606 & 0.26\\
        &       &      & 34.5 $\rightarrow$ 108.3 min & CR=16.0 $\alpha$=1.23 & \citet{Klotz2005} & c & 0.606 & 0.26\\
050730 & SWIFT & 35s & 66 - 124 s & CR=16.7 $\pm$ 0.4 & GCNC 3720  & c & 3.967 & 0.14\\
       &       &     & 23.8 $\rightarrow$ 86.4 min & CR=16.8  $\alpha$=0.39 & This study  & c & 3.967 & 0.14\\
050803 & SWIFT & 110s & 61 - 63 min & CR$>$16.0 & GCNC 3751  & c & & 0.20\\
		 &       &      & 84 - 156 min & CR$>$18.7 & GCNC 3751  & c & & 0.20\\
050824 & SWIFT & 70s & 9.8 - 10.1 min & CR$>$16.4 & This study  & c & 0.83 & 0.09\\
       &       &         & 33.0 - 48.8 min & CR=19.3 $\pm$ 0.4 & This study  & c & 0.83 & 0.09\\
050904 & SWIFT & 225s & 86 - 144 s$^*$ & CI$>$15.3 & \citet{Boer2006}  & c & 6.29 & 0.17\\
       &       &    & 150 - 253 s$^*$ & CI=15.2 $\pm$ 0.3& \citet{Boer2006}  & c & 6.29 & 0.17\\
       &      & & 312 - 443 s & CI=15.3 $\pm$ 0.3 & \citet{Boer2006}  & c & 6.29 & 0.17\\
       &      & & 449 - 589 s & CI=14.1 $\pm$ 0.3 & \citet{Boer2006}  & c & 6.29 & 0.17\\
       &      & & 595 - 978 s & CI$\simeq$15.5 $\pm$ 0.4 & \citet{Boer2006} & c & 6.29 & 0.17\\
       &      & & 985 - 1666 s & CI$>$15.8 & \citet{Boer2006}  & c & 6.29 & 0.17\\
051211A & HETE & $<$1s & 3.14 - 3.19 h & CR$>$15.3 & GCNC 4329  & c & {\it 4.9 $\pm$ 2.0} & 0.32 \\
051211B & INTEGRAL & 80s & 131 - 161 s & CR$>$16.2 & GCNC 4328  & c & & 1.25\\
051221B & SWIFT & 192s & 216 $\rightarrow$ 276 s & CR$>$14 & GCNC 4386  & c & & 3.71\\
        &       &      & 282 - 312 s           & CR$>$18.2 & GCNC 4386  & c & & 3.71\\
060111A & SWIFT & 20s  & 20 $\rightarrow$ 80 s & CR$>$16.2 & GCNC 4483  & c & & 0.08\\
        &       &      & 86 - 261 s & CR$=$18.3 $\pm$ 0.4 & GCNC 4483  & c & & 0.08\\
        &       &      & 510 - 1027 s & CR$>$18.7 & GCNC 4483  & c & & 0.08\\
060111B & SWIFT & 59s & 28 $\rightarrow$ 88 s$^*$ & CR=13.8 $\alpha$=2.38  & \citet{Klotz2006}  & c & & 0.30\\
        &       &      & 94 $\rightarrow$ 198 s & CR=16.75 $\alpha$=1.08  & \citet{Klotz2006}  & c & & 0.30\\
060124 & SWIFT & 710s & 1.98 - 2.71 h & CR$<$18.3 & This study & c & 2.297 & 0.36\\
060502B & SWIFT & $<$1s & 2.08 - 2.60 h & CR$>$16.8 & This study & c & 0.287? & 0.12\\
060512 & SWIFT & 8.6s & 13.3 - 34.7 min & CR=17.6 $\pm$ 0.3 & GCNC 5140 & c & 0.4428 & 0.05\\
       &       &      & 34.8 - 228.3 min & CR=19.0 $\pm$ 0.6 & GCNC 5140 & c & 0.4428 & 0.05 \\
060515 & SWIFT & 52s & 60 $\rightarrow$ 120 s & CR$>$15.1 & GCNC 5134 & c & & 0.08\\
       &       &  & 126 - 156 s & CR$>$16.2 & GCNC 5134 & c & & 0.08 \\
       &       &  & 2.1 - 25.0 min & CR$>$18.2 & GCNC 5134 & c & & 0.08 \\
060712 & SWIFT & 26s & 104 $\rightarrow$ 164 s & CR$>$12.5 & This study & c & & 0.03\\
       &       &  & 171 - 458 s & CR$>$13.8 &  This study & c & & 0.03\\
060814 & SWIFT & 140s & 21 $\rightarrow$ 81 s$^*$ & CR$>$14.8 & GCNC 5448 & c & 0.84 & 0.11\\
       &       &   & 88 - 118 s$^*$ & CR$>$15.7 & GCNC 5448 & c & 0.84 & 0.11\\
       &       &   & 125 - 155 s$^*$ & CR$>$15.7 & GCNC 5448 & c & 0.84 & 0.11\\
       &       &   & 161 - 191 s & CR$>$15.7 & GCNC 5448 & c & 0.84 & 0.11\\
       &       &   & 88 - 228 s & CR$>$17.4 & GCNC 5448 & c & 0.84 & 0.11\\
060904A & SWIFT & 80s & 27 $\rightarrow$ 87 s$^*$ & CR$>$16.0 & GCNC 5506 & c & {\it 1.84 $\pm$ 0.85} & 0.05 \\
       &        &  & 94 - 124 s & CR$>$17.3 & GCNC 5506 & c & {\it 1.84 $\pm$ 0.85} & 0.05 \\
       &        & & 94 - 376 s & CR$>$19.5 & GCNC 5506 & c & {\it 1.84 $\pm$ 0.85} & 0.05 \\
060904B & SWIFT & 192s &   23 $\rightarrow$ 41 s$^*$ & CR$>$17.2  & \citet{Klotz2008} & c & 0.703 & 0.46 \\
       &        & &  41 $\rightarrow$ 44 s$^*$     & CR$>$15.9  & \citet{Klotz2008} & c & 0.703 & 0.46 \\
       &        &    &  44 $\rightarrow$ 86 s$^*$     & CR$>$17.2  & \citet{Klotz2008} & c & 0.703 & 0.46 \\
       &        &    & 1.50 $\rightarrow$ 9.45 min$^*$ & CR=18.16 $\alpha$=-0.80 & \citet{Klotz2008} & c & 0.703 & 0.46 \\
       &        &    & 9.45 $\rightarrow$ 21.7 min & CR=16.57 $\alpha$=1.38 & \citet{Klotz2008} & c & 0.703 & 0.46 \\
       &        &    & 21.7 $\rightarrow$ 41.8 min & CR=17.8 $\pm$ 0.1 $\alpha$=0 & \citet{Klotz2008} & c & 0.703 & 0.46 \\
060919 & SWIFT & 9s & 28 $\rightarrow$ 88 s & CR$>$15.4 & GCNC 5576 & s & & 0.19\\
       &       &  & 94 - 124 s & CR$>$15.8 & GCNC 5576 & s & & 0.19 \\
       &       &  & 94 - 196 s & CR$>$15.9 & GCNC 5576 & s & & 0.19 \\
061019 & SWIFT & 191s & 15.0 - 16.5 min & CR$>$14.8 & GCNC 5731 & s & & 3.01 \\
       &       &  & 15.0 - 23.2 min & CR$>$15.8 & GCNC 5731 & s & & 3.01 \\
061027 & SWIFT & 170s & 14.0 - 18.6 h & R$>$20.0 & GCNC 5788 & s & & 0.41 \\
       &       &  & 14.0 - 18.6 h & I$>$18.9 & GCNC 5788 & s & & 0.41 \\
061028 & SWIFT & 106s & 145 $\rightarrow$ 205 s & CR$>$16.5 & GCNC 5763 & c & & 0.42\\
       &       &  & 212 - 242 s & CR$>$17.5 & GCNC 5763 & c & & 0.42 \\
       &       &  & 212 - 497 s & CR$>$18.3 & GCNC 5763 & c & & 0.42 \\
       &       &  & 0.44 - 1.07 h & CR$>$20.1 & GCNC 5764 & c & & 0.42 \\
061110B & SWIFT & 128s & 10.4 - 11.9 min & CR$>$16.6 & GCNC 5801 & c & 3.44 & 0.11\\
       &        & & 10.4 - 18.3 min & CR$>$16.9 & GCNC 5801 & c & 3.44 & 0.11\\
061217 & SWIFT & 0.3s & 20.2 $\rightarrow$ 80.2 s & CR$>$16.5 & GCNC 5942 & c & 0.827 & 0.12\\
       &       &   & 87 - 117 s & CR$>$18.1 & GCNC 5942 & c & 0.827 & 0.12\\
       &       &       & 87 - 371 s & CR$>$19.2 & GCNC 5942 & c & 0.827 & 0.12\\
061218 & SWIFT &  4s  & 61 $\rightarrow$ 121 s & CR$>$15.8 & GCNC 5980 & s & & 0.34\\
       &       &      & 127 - 157 s & CR$>$17.4 & GCNC 5943 & s & & 0.34\\
       &       &      & 162 - 192 s & CR$>$17.4 & GCNC 5943 & s & & 0.34\\
       &       &      & 127 - 511 s & CR$>$18.0 & GCNC 5943 & s & & 0.34\\
061222B & SWIFT & 40s & 84 $\rightarrow$ 108 s & CR$>$17.7 & GCNC 5981 & s & 3.355 & 1.01\\
       &       &    & 108 $\rightarrow$ 144 s & CR$>$14.7 & GCNC 5981 & s & 3.355 & 1.01\\
       &       &    & 150 - 180 s & CR$=$18.2 $\pm$ 0.5 & GCNC 5960 & s & 3.355 & 1.01\\
       &       &    & 185 - 215 s & CR$>$18.2 & GCNC 5981 & s & 3.355 & 1.01\\
       &       &    & 150 - 534 s & CR$>$19.9 & GCNC 5959 & s & 3.355 & 1.01\\
070103 & SWIFT & 19s & 23 $\rightarrow$ 83 s & CR$>$15.1 & GCNC 5989 & c & & 0.18\\
       &       &      & 89 - 120 s & CR$>$15.9 & GCNC 5989 & c & & 0.18\\
       &       &      & 89 - 229 s & CR$>$16.4 & GCNC 5989 & c & & 0.18\\
070227 & SWIFT & 7s  & 7.8 - 9.0 h & CR$>$20.2 & GCNC 6159 & s & & 0.90\\
070330 & SWIFT & 9s  & 2.5 - 3.5 h & CR$>$17.4 & GCNC 6235 & s & & 0.17\\
070411 & SWIFT & 101s & 46 $\rightarrow$ 106 s$^*$ & CR$>$14.6 & GCNC 6487 & c & 2.954 & 0.76\\
       &       &    & 113 - 143 s & CR$>$15.5 & GCNC 6487 & c & 2.954 & 0.76\\
070412 & SWIFT & 34s  & 11.3 - 12.8 min & CR$>$18.5 & GCNC 6488 & c & & 0.06\\
       &       &       & 11.3 - 20.9 min & CR$>$19.5 & GCNC 6488 & c & & 0.06\\
070420 & SWIFT & 77s          & 40 $\rightarrow$ 49 s$^*$ & CR$>$16.75 &  \citet{Klotz2008} & s & {\it 1.56 $\pm$ 0.35}& 1.39\\
       &       & & 49 $\rightarrow$ 57 s$^*$ & CR$>$16.45 &  \citet{Klotz2008} & s & {\it 1.56 $\pm$ 0.35} & 1.39\\
       &       &               & 57 $\rightarrow$ 66 s$^*$ & CR$>$16.25 &  \citet{Klotz2008} & s & {\it 1.56 $\pm$ 0.35} & 1.39\\
       &       &               & 66 $\rightarrow$ 91 s$^*$ & CR$>$16.75 &  \citet{Klotz2008} & s & {\it 1.56 $\pm$ 0.35} & 1.39\\
       &       &               & 116 $\rightarrow$ 188 s & CR=16.7 $\alpha$=-1.26 & \citet{Klotz2008} & s  & {\it 1.56 $\pm$ 0.35} & 1.39\\
       &       &               & 341 $\rightarrow$ 1031 s & CR=16.4 $\alpha$=0.88 & \citet{Klotz2008} & s  & {\it 1.56 $\pm$ 0.35} & 1.39\\
070508 & SWIFT & 40s  & 13.8 - 15.3 min & CR$>$17.8 & GCNC 6384 & s & & 0.37\\
       &       &       & 13.8 - 28.4 min & CR$>$19.7 & GCNC 6384 & s & & 0.37\\
070521 & SWIFT & 38s  & 24 $\rightarrow$ 84 s$^*$ & CR$>$17.1 &  GCNC 6434 & s & {\it 2.28 $\pm$ 0.45} & 0.07\\
       &       & & 89 - 119 s & CR$>$17.5 & GCNC 6434 & s & {\it 2.28 $\pm$ 0.45} & 0.07\\
       &       &               & 89 - 570 s & CR$>$19.0 & GCNC 6434 & s & {\it 2.28 $\pm$ 0.45} & 0.07\\
070621 & SWIFT & 33s  & 23 $\rightarrow$ 83 s$^*$ & CR$>$14.9 &  GCNC 6563 & c & & 0.13\\
       &       &       & 90 - 120 s & CR$>$15.4 &  GCNC 6563 & c & & 0.13\\
070628 & SWIFT & 39s  & 8.3 - 9.1 h & CR$>$17.9 &  GCNC 6588 & s & & 2.42\\
070913 & SWIFT & 8s  & 87 - 117 s & CR=17.5 $\pm$ 0.5 &  GCNC 6787 & s & & 0.35\\
       &       &      & 124 - 154 s & CR=18 $\pm$ 0.5 &  GCNC 6787 & s & & 0.35\\
       &       &      & 288 - 378 s & CR$>$18 &  GCNC 6787 & s & & 0.35\\
070920A & SWIFT & 56s  & 9.1 - 10.6 min & CR$>$18.2 &  GCNC 6806 & c & & 0.29\\
070920B & SWIFT & 20s  & 2.7 - 3.6 h & CR$>$18.2 &  GCNC 6816 & s & & 0.04\\
071010A & SWIFT & 6s & 124 $\rightarrow$ 514 s & CR=17.5 $\alpha$=-0.89 & This study & s & 0.98 & 0.28\\
        &       &    & 8.8 $\rightarrow$ 88 min & CR=16.12 $\alpha$=0.64 & This study & s & 0.98 & 0.28\\
071101 & SWIFT & 3s   & 39 $\rightarrow$ 99 s & CR$>$16.3 &  GCNC 7035 & c & & 3.99\\
       &       &       & 106 - 136 s & CR$>$17.5 &  GCNC 7035 & c & & 3.99\\
       &       &       & 106 - 394 s & CR$>$18.4 &  GCNC 7035 & c & & 3.99\\
071109 & INTEGRAL & 30s  & 36 - 96 s & CR$>$15.4 &  GCNC 7052 & c & & 1.74\\
       &       &         & 103 - 133 s & CR$>$15.5 &  GCNC 7052 & c & & 1.74\\
071112B & SWIFT & 0.30s  & 5.8 - 10.5 h & R$>$20.4 &  This study & s & & 0.56\\
071112C & SWIFT & 15s  & 1.6 - 2.6 h & CR=20.1 &  GCNC 7065 & c & 0.8230 & 0.32\\
071117 & SWIFT & 6.6s  & 9.5 - 11.3 h & CI$>$20 &  GCNC 7108 & s &  & 0.06\\
080129 & SWIFT & 80s  & 59 - 74 min & CR$>$16.4 & GCNC 7234 & s & & 2.724\\
080207 & SWIFT & 300s  & 26 - 27 min & CR$>$14.3 & GCNC 7267 & c & & 0.062\\
080210 & SWIFT & 45s & 390 $\rightarrow$ 3162 s & CR=16.62 $\alpha$=1.21 & This study & s & 2.64 & 0.223\\
080315 & SWIFT & 45s  & 85 - 91 s & CR$>$16.8 & GCNC 7417 & c & & 0.036\\
080330 & SWIFT & 61s & 20.4 $\rightarrow$ 80.4 s & CR$>$16.6 & This study & c & 1.51 & 0.044\\
       &       &     & 121 $\rightarrow$ 333 s & CR=17.6 $\alpha$=-0.8  & This study & c & 1.51 & 0.044\\
       &       &     & 428 $\rightarrow$ 1012 s & CR=16.75 $\alpha$=0.5  & This study & c & 1.51 & 0.044\\
080413A & SWIFT & 46s  & 480 - 520 s & CR$>$12.5 & GCNC 7595 & c & 2.433 & 0.423\\
        &       &      & 1174 - 1263 s & CR=15.0 $\pm$ 0.2 & GCNC 7595 & c & 2.433 & 0.423\\
        &       &      & 1470 - 1650 s & CR=15.3 $\pm$ 0.2 & GCNC 7595 & c & 2.433 & 0.423\\
        &       &      & 2035 - 2215 s & CR=15.5 $\pm$ 0.2 & GCNC 7595 & c & 2.433 & 0.423\\
080430 & SWIFT & 16s & 18.8 $\rightarrow$ 25.2 s$^*$ & CR$>$16.7 & This study & c & 0.767 & 0.033\\
       &       &     & 31 $\rightarrow$ 41 s & CR=16.65 $\alpha$=-3.53  & This study & c & 0.767 & 0.033\\
       &       &     & 48 $\rightarrow$ 1250 s & CR=15.54 $\alpha$=0.65  & This study & c & 0.767 & 0.033\\
080516 & SWIFT & 5.8s & 65 $\rightarrow$ 124 s & CR$>$13.5 & GCNC 7798 & s & {\it 3.2 $\pm$ 0.3} & 1.1\\
       &       &     & 131 - 161 s & CR$>$15   & GCNC 7798 & s & {\it 3.2 $\pm$ 0.3} & 1.1\\
080603B & SWIFT & 70s & 1.03 $\rightarrow$ 4.7 h & CR=17.3 $\alpha$=0.62 & This study & c & 2.69 & 0.05\\
080903 & SWIFT & 66s & 29.1 $\rightarrow$ 89.1 s$^*$ & CR$>$17.1 & This study & c &  & 0.556\\
       &       &     & 96 - 126 s & CR$>$18.2 & This study & c &  & 0.556\\
\enddata
\end{deluxetable}

\end{document}